\documentstyle{elsart}
\begin{document}
\begin{frontmatter}
\title{Scale-free properties of weighted networks with connectivity-driven 
topology}
\author{W. Je\.zewski\thanksref{X}}
\address{Institute of Molecular Physics, Polish Academy of Sciences, 
Smoluchowskiego 17, 60-179 Pozna\'n, Poland}
\thanks[X]{e-mail: jezewski@ifmpan.poznan.pl}
\begin{abstract}
The rate equations are used to study the scale-free behavior of the weight 
distribution in evolving networks whose topology is determined only by degrees 
of preexisting vertices. An analysis of these equations shows that the degree 
distribution and thereby the weight distribution remain unchanged when the 
probability rate of attaching new nodes is replaced with an unnormalized rate 
determined by the ratio of the degree of a randomly selected old node to the 
maximal node degree at the current stage of the network evolution. Such a 
modification of the attachment rule is argued to accelerate considerably 
numerical simulations of both unweighted and weighted networks belonging to 
the class of investigated evolving systems. It is also proved that the studied 
rate equations have a solution corresponding to the total weight (concentrated 
at individual vertices) distribution displaying the power-law behavior for 
asymptotically large weights.
\end{abstract}

\end{frontmatter}

\section{Introduction}

Scale-free networks belong to an important group of complex evolving systems 
exhibiting a power-law of the degree (connectivity) distribution \cite{al}. To 
explain the origin of the scaling form of the degree distribution, several 
scenarios of the growth and self-organization of networks have been proposed 
\cite{b1,b2,b3,hu,go,d1,d2,d3,va,kl}. Probably the most natural 
mechanism responsible for scaling properties of time-evolving networks is the 
preferential attachment of a new vertex with the probability being 
proportional to the degree of a randomly selected preexisting node to which a 
new one will be connected \cite{b1,b2,b3}. This simple self-organization 
mechanism allows one to recover the complex topological structure of scale-free
networks \cite{al,b2}. However, in many real network systems, links perform 
some specific functions, and the underlying binary networks can only be 
considered as skeletons determining topological properties of these systems. 
Thus, in real networks, links play a role of arteries through which some 
processes of different intensities are executed, or which carry loads of different 
magnitudes \cite{n1,n2,ma,r1,r2,r3}. Functional properties of networks can be 
characterized by assigning loads or weights to particular links \cite{yo}. 
Then, the network evolution is in general described by both preferential node 
attaching and assigning weights to links. Essentially, processes of the 
preferential node attachment and the link loading are not autonomous. The 
preferential mechanism of the vertex attachment can be determined only by 
degrees of existing vertices \cite{yo}, but can also be affected by weights, 
or can even be controlled entirely by total weights concentrated at 
preexisting vertices \cite{r1,r2,r3}. On the other hand, weights ascribed to 
newly created links can depend on degrees of old nodes to which new nodes are 
connected \cite{yo}, or, in cases when internal (existing) links are loaded, 
the weight assigned to a given internal link can be expressed by degrees of 
both vertices joined by this link \cite{n1}. Models in which the attachment 
rule is entirely governed by degrees of existing nodes appear to be 
appropriate for describing not only topological but also functional features 
of the WWW. In this network, the decision about linking a pair of nodes 
(webpages) depends on the popularity (measured by the node degree) of a given 
preexisting node to which an another node will be connected rather than on 
the weight concentrated at the existing node, i.e., the total size of 
documents sent from this node to others. Numerical simulations of networks 
with degree-driven topologies indicate that the total weight distribution 
can display in such systems a power-law form \cite{yo,j1}. In turn, models 
whose topologies are entirely controlled by weights provide a representation 
of real systems in which a change of network traffic due to connecting a 
new node to an old node strongly disturbs functions performed by all links 
joining the old node with other nodes, or in which connections of new 
vertices influence functions of whole systems. Illustrative examples of such 
real networks are transportation systems \cite{r1,r2,r3}. It has been shown 
analytically that, in the case of linear dependence of the attachment 
probability on total weights concentrated at vertices, the distribution of 
total weights can also display a power-law form \cite{r1,r2,r3}.

In this paper, undirected weighted networks with connectivity driven 
topologies are considered. By analyzing the rate equations for the degree and 
weight distributions in such networks, it is shown that these distributions 
remain unchanged as the preferential mechanism of node attaching is 
appropriately modified by using an unnormalized attachment rule. This change 
of the vertex attaching rule is shown to be very useful for numerical 
simulations of some unweighted (binary) networks as well as for simulations 
of the considered weighted networks with degree-governed topological 
structures. The rate equations for the time evolution of the distribution of 
total weights concentrated at vertices of a given connectivity is proved to 
have a solution of a scale-free form.    

\section{Evolution of the degree distribution in binary networks}

In this section, the preferential attachment mechanism that a new node is 
linked to a randomly chosen preexisting node with the probability depending on 
the degree of the earlier node is revised. In general, this probability is 
assumed in the form \cite{b3,k1} 
\begin{equation}
\mit{\Pi}_\gamma(k)=A_\gamma(k)/M_\gamma(t)
\end{equation}
with the connection kernel
\begin{equation}
A_\gamma(k)=
\left\{
\begin{array}{ll}
k^\gamma\,\,\,&,\,\,\,\gamma\ne 1\,\,\,,\\
a_kk\,\,\,&,\,\,\,\gamma=1\,\,\,,
\end{array}
\right.
\end{equation}
and with the normalization factor $1/M_\gamma(t)$ determined by
\begin{equation}
M_\gamma(t)=\sum_{j=1}^{k_{\rm max}} A_\gamma(j)\,N_j(\gamma;t)\,\,\,,\\
\end{equation}
where $\gamma\ge 0$ is the nonlinearity exponent, $k$ denotes the vertex 
connectivity, $k_{\rm max}=k_{\rm max}(\gamma;t)$ is the maximal 
node degree, the amplitudes $a_k\ge 0$, $k=1,2,...$, and $N_k(\gamma;t)$ 
denotes the number of vertices of the degree $k$. Note that the power-law 
$N_k\sim k^{-\nu}$ emerges only if the kernel $A_\gamma(k)$ is linear or 
asymptotically linear, i.e., when $\gamma=1$ and the amplitudes $a_k$ are 
all equal, or are unequal, but such that $a_k\rightarrow a_\infty$ as 
$k\rightarrow\infty$ \cite{k1}.

It is proved below that the degree distribution remains unchanged when the 
probability attachment rate $\mit{\Pi}_\gamma(k)$ is replaced with the rate
\begin{equation}
\tilde{\mit{\Pi}}_\gamma(k)=
\left\{
\begin{array}{ll}
k^\gamma/k_{\rm max}^\gamma(\gamma;t)\,\,\,&,\,\,\,\gamma\ne 1\,\,\,,\\
a_kk/a_\infty k_{\rm max}(1;t)\,\,\,&,\,\,\,\gamma=1\,\,\,.
\end{array}
\right.
\end{equation}
For cases of $0\le\gamma\le 1$, this modified attachment rate is unnormalized 
(thus, it does have the character of a probability rate), and is not proportional 
to the rate ${\mit{\Pi}}_\gamma(k)$ as, in these cases, time dependences of 
$k_{\rm max}(\gamma;t)$ and $M_\gamma(t)$ are different. If $a_k>a_\infty$ 
(for some $k<k_{\rm max}$), the rate $\tilde{\mit{\Pi}}_1(k)$ takes even values 
greater than one. To show the equivalence of both preferential linking 
procedures, one can apply the rate equations describing, for the attachment 
rate (1), the time evolution of the degree distribution \cite{k1}
\begin{equation}
\frac{d}{dt}N_k(\gamma;t)=\frac{1}{M_\gamma(t)}[\,A_\gamma(k-1)\,
N_{k-1}(\gamma;t)-A_\gamma(k)\,N_k(\gamma;t)\,]+\delta_{k,1}\,\,\,.
\end{equation}
Solutions of these equations imply for large $t$ that \cite{k1,k2}
\begin{equation}
M_\gamma(t)=
\left\{
\begin{array}{ll}
\mu t\,\,\,&,\,\,\,0\le\gamma\le1\,\,\,,\\
t^\gamma\,\,\,&,\,\,\,\gamma>1\,\,\,,
\end{array}
\right.
\end{equation}
where $\mu$ is a function of $\gamma$, such that, in cases of homogeneous 
kernels with uniform, unit amplitudes, $\mu\in[1,2]$ for all $\gamma\ge 0$ 
($\mu=1$ for $\gamma>1$), and
\begin{equation}
k_{\rm max}(\gamma;t)=
\left\{
\begin{array}{ll}
(\frac{1-\gamma}{\mu})^{\frac{1}{1-\gamma}}\,(\ln\,t)^{\frac{1}{1-\gamma}}
\,\,\,&,\,\,\,0\le \gamma <1\,\,\,,\\
s(\nu)\,t^{\frac{1}{\nu-1}}\,\,\,&,\,\,\,\gamma=1\,\,\,,\\
t\,\,\,&,\,\,\,\gamma>1\,\,\,,
\end{array}
\right.
\end{equation}
with $\nu\ge 2$, $s$ being dependent on $\nu$ as well as on the manner of the 
convergence of $a_k$ to $a_\infty$ as $k\rightarrow\infty$ (in the case of 
$\gamma$=1). Since, $k_{\rm max}(\gamma;t)$ is a monotonic function of time 
for each $\gamma$, Eq. (5) can be transformed from $t$ to $k_{\rm max}$ using 
the inverse function $t=g_\gamma(k_{\rm max})$. Thus, the use of Eqs. (6) 
and (7) gives
\begin{equation}
M_{\gamma}(t)=k_{\rm max}^{\gamma}\,g'_{\gamma}(k_{\rm max})\,\,\,,
\end{equation}
where, $g'_\gamma$ denotes the derivative of $g_\gamma$ with respect to 
$k_{\rm max}$. Next, inserting (8) into Eq. (5), replacing $t$ by $k_{\rm max}$, 
and multiplying these equations by $g'_\gamma(k_{\rm max})$, one obtains
\begin{eqnarray}
\!\!&&\frac{d}{dk_{\rm max}}\tilde{N}_k(\gamma;k_{\rm max})= \nonumber \\
\!\!&&\frac{1}{k_{\rm max}^\gamma}[\,\tilde{A}_\gamma(k-1)\,
\tilde{N}_{k-1}(\gamma;k_{\rm max})-\tilde{A}_\gamma(k)\,
\tilde{N}_k(\gamma;k_{\rm max})\,]+ \delta_{k,1}\,g'_\gamma(k_{\rm max})
\end{eqnarray}
with $\tilde{N}_k(\gamma;k_{\rm max})\equiv N_k(\gamma;g_\gamma(k_{\rm max}))$,
$\tilde{A}_\gamma(k)\equiv A_\gamma(k)$ for $\gamma\ne 1$ and 
$\tilde{A}_1(k)=(a_k/a_\infty)A_1(k)$. In Eq. (9), the role of time is played 
by the variable $k_{\rm max}$. The factor $g'_{\gamma}$ in the last term on the right 
hand side of (9) determines for $0\le\gamma\le 1$ the acceleration of introducing new 
nodes (of degree one), as $k_{\rm max}$ increases (the number of new nodes introduced 
in successive equal intervals of $k_{\rm max}$ is for $0\le\gamma\le 1$ a growing 
function of $k_{\rm max}$). Note that $g'_\gamma(k_{\rm max})$=1 for $\gamma>1$. The 
rate equation (9) can be solved using the substitutions (cf. Ref. 20) 
$\tilde{N}_k(\gamma;k_{\rm max})=n_kg_\gamma(k_{\rm max})$ for $0\le\gamma\le 1$ and 
$\tilde{N}_k(\gamma;k_{\rm max})=J_kg_\gamma^{k-(k-1)\gamma}(k_{\rm max})$ for 
$\gamma>1$. Thus, according to (7), the systems of equations (5) and (9) are 
equivalent and, consequently, solutions to them yield the same degree distribution. 
However, instead of the original attachment rate, determined by Eqs. (1)-(3), the rate 
occurring in Eq. (9) is of the form (4). These rates satisfy for $\gamma\le 1$ the 
inequality $\tilde{\mit{\Pi}}_\gamma(k)<\mit{\Pi}_\gamma(k)$, $k=1,2,...$, except for 
cases of some $a_k<a_\infty$ (when $\gamma=1$), for which the inequality holds only 
for sufficiently large $k$. Therefore, using the preferential rate (4) in 
numerical simulations of evolving binary networks enables one to speed up 
computations, with regard to analogous computations based on employing the 
rate (1), for the same exponent $\gamma\le 1$ and for the same set of amplitudes 
$\{a_k\}$ (when $\gamma=1$). The speed up ratio of computations at a given stage of 
the network generation (i.e. at a given moment $t$) is determined by the quotient of 
the attachment probabilities (4) and (1), or by the quotient of $M_\gamma(t)$ and 
$k_{\rm max}^\gamma(\gamma;t)$. By means of (8), this momentary ratio is given by 
$\tilde{\tau}_\gamma(t)=g'_\gamma(k_{\rm max}(\gamma;t))$. Clearly, 
$\tilde{\tau}_\gamma(t)$ is equal to the evolution acceleration factor occurring in 
the last term on the right hand side of Eq. (9). The total speeding up ratio 
$\tau_\gamma(N)$, determining the reduction of the entire period of time during 
which a network is numerically generated, beginning with one node at the initial 
time $t_i=1$ and ending with $N$ nodes at the final time $t_f=N$, can be estimated 
using the relation
\begin{equation}
\tau_\gamma(N)=(N-1)^{-1}\!\!\!\!\int_{\hspace{3mm} 1}^{\hspace{6mm} N}
\!\!g'_\gamma(k_{\rm max}(\gamma;t))\,dt\,\,\,.
\end{equation}   
By applying (7), one finds for $0\le\gamma\le 1$ and for large $N$ that 
\begin{equation}
\tau_\gamma(N)\sim g'_\gamma(k_{\rm max}(\gamma;N))\,\,\,,
\end{equation}
and hence
\begin{equation}
\tau_\gamma(N)\sim
\left\{
\begin{array}{ll}
(\ln N)^{-\frac{\gamma}{1-\gamma}}N\,\,\,&,\,\,\,0\le\gamma<1\,\,\,,\\
N^{\frac{\nu-2}{\nu-1}}\,\,\,&,\,\,\,\gamma=1\,\,\,.
\end{array}
\right.
\end{equation}
Notice that, in the case of $\gamma>1$, Eqs. (10) and (7) imply that 
$\tau_\gamma(N)=1$. Because $\tau_\gamma$ is an increasing function of $N$ for each 
$\gamma$ from the interval $[0,1]$, the application of the rate (4) in numerical 
calculations in cases of $0\le\gamma\le 1$ becomes more and more effective as 
$N$ grows (compared with analogous calculations with the use of the rate (1)). 
Clearly, this is of rather little importance, as Eq. (5) are exactly solvable 
for all $k$ when $\gamma=1$ and for asymptotically large $k$ when 
$0\le\gamma\le 1$. It turns out, however, that, in simulations of some networks 
with more complex normalized linking rules than given by Eq. (1), the 
replacement of normalized attaching rate with an unnormalized rate of the 
type (4) is also possible \cite{j2}. Furthermore, the application of the linking 
rate (4) can be very useful for numerical simulations of weighted networks with 
connectivity-driven topologies, especially in cases when the weight distribution 
cannot be found exactly. 

Time needed to simulate networks evolved by node attaching with the 
probability rate with the strictly linear kernel can also be reduced by 
using, instead of the attachment rate (1), a uniform initial attachment 
rate  (corresponding to $\gamma=0$), combined with the redirection, with  
probability $r=0.5$, of newly created links to ancestors of selected 
preexisting nodes \cite{k2}. Another way to enhance the efficiency of 
simulations of growing networks is the application of parallel 
algorithms \cite{mc}. Although by using different evolution rules one can 
generate in the large $N$ limit networks revealing the same (at least for large 
connectivities) degree distribution, these networks can exhibit different 
histories of growth. However, the statistics of weights introduced in such 
connectivity-driven networks cannot necessarily be independent of histories of 
the network growth. This can happen when loads are ascribed to links as soon as 
the links are created, and when loads magnitudes depend strongly on details 
of network histories. Consequently, weighted networks can display different 
weight distributions, even if they undergo the same rule of link loading and 
even if they exhibit the same degree distribution. Therefore, the rate equation 
approach to study the weight distribution in networks with connectivity-governed 
topologies should, in general, incorporate rate equations for the evolution of 
the degree distribution.  

\section{Evolution of weighted networks}

The process of link loading can be carried out in many different ways, 
according to various functions performed by real networks \cite{r1,r2,r3,yo,j1}. 
For example, to mimic functional properties of transportation or communication 
networks, an evolving model has been elaborated by assuming that the topology 
of the underlying skeleton network is entirely determined by total weights 
concentrated at vertices to which new vertices are connected, and by assuming 
that the procedure of ascribing a weight to a newly created link involves a 
redistribution (perturbation) of weights, assigned to all links concentrated at 
the existing vertex to which new vertex has been connected \cite{r1,r2,r3}. 
However, there exist also real networks whose topologies are governed by vertex 
degrees only. Perhaps the most representative example of such networks is the 
WWW. The appearance of a new link in this network depends on the popularity of 
a given node (webpage) that connects a new node rather than depends on the node 
weight (the total size of documents taken from the preexisting webpage). 
Naturally, the popularity of a given webpage can be measured by its 
connectivity (the number of directed hyperlinks joining this webpage with other 
webpages). Although traffic jams in the WWW reduce the speed of sending 
documents from one webpage to others, the total size (usually unknown) of 
documents sent from a given webpage has no effect on the mere act of creating a 
hyperlink between this webpage and another one. Thus, the topology of the WWW 
can be considered as independent of weights assigned to links. 

Here, weighted networks in which node attaching is determined by the 
probability rate (1) (or by the rate (4)) are studied. For simplicity, only 
the linear case $\gamma=1$, $a_k=1$, $k=1,2,...$, of the preference rate is 
discussed. Each of the links $i\!\!\leftrightarrow\!\! j$ of a given network 
is taken to be loaded by a weight $w_{i,j}>0$, and, consequently, each vertex 
is assumed to concentrate a total weight $w_i=\sum_j w_{i,j}$. The number of 
nodes with $k$ links and the total weight $w$ is given by
\begin{equation}
N_k(w,t)=\sum_{i=1}^{N_k(t)}{\rm \delta}(w-w_i)\,\,\,,
\end{equation}
where the sum runs over all $N_k(t)$ nodes with $k$ links and ${\rm \delta}$ 
denotes the Dirac delta function. The rate equations for the evolution of 
the number $N_k(w,t)$ of nodes with $k$ links and with the total weight $w$ 
can be expressed in the form 
\begin{eqnarray}
\frac{\partial}{\partial t}& &N_k(w,t)= \nonumber \\
& &\frac{k-1}{M(t)}\hspace{-12mm}\int_{\hspace{7mm}\varepsilon_{k-1}}^
{\hspace{16mm} w-E_{k-1}}
\rho_{k-1}(x)\,N_{k-1}(w-x,t)\,dx-\frac{k}{M(t)}\,N_k(w,t)+ \nonumber \\
& &\delta_{k,1}\,\rho_k(w-E_k+\varepsilon_k)\,\,\,,
\end{eqnarray}
where $M(t)\equiv M_1(t)=\sum_{j}jN_j(t)$, $\varepsilon_{k-1}$ is the minimal 
weight that can be introduced to any vertex with $k-1$ links in consequence 
of connecting  a new link, $E_{k-1}$ denotes a minimal of total weights 
concentrated at vertices of the degree $k-1$, and $\rho_k(x)$ is the 
probability distribution of weights that can be assigned to a new link 
connected to any of vertices of the degree $k$. Similarly to the limit 
weights $\varepsilon_k$ and $E_k$, the distribution $\rho_k(x)$  is determined 
by rules which govern the weight loading \cite{j1}. The first term on the right 
hand side of Eq. (14) is responsible for different processes of the increase 
of total node weights from the value $w-x$ to $w$, as a result of connecting 
new links with weights $x\in[\varepsilon_{k-1},w-E_{k-1}]$ (note that 
$w\ge E_{k-1}+\varepsilon_{k-1}=E_k$). The second and third terms in (14) are 
the loss and the new site terms, respectively \cite{k1}. Obviously, integrating 
Eq. (14) over all $w$ one recovers the rate equation (5) for the connectivity 
distribution, in the case of the linear connection kernel. In accordance with 
the argumentation presented in the previous section, the evolution of 
$N_k(w,t)$ can alternatively be investigated with respect to $k_{\rm max}$. 

Since $M(t)=2t$ \cite{k1}, one can substitute $N_k(w,t)=tn_k(w)$. Hence, using 
(14), one obtains
\begin{eqnarray}
n_k(w)&=&\frac{k-1}{k+2}\,\hspace{-12mm}
\int_{\hspace{7mm}\varepsilon_{k-1}}^{\hspace{16mm} w-E_{k-1}}\rho_{k-1}(x)\,
n_{k-1}(w-x)\,dx+ \nonumber \\
& &\frac{2}{k+2}\,\delta_{k,1}\,\rho_k(w-E_k+\varepsilon_k)\,\,\,.
\end{eqnarray}

To show that there exists for large $w\gg \varepsilon_k,E_k$ a solution to 
Eq. (15) of the scaled form $n_k(w)\sim w^{-\alpha}$, $\alpha>1$, assume that 
\begin{equation}
n_k(w)\simeq n_kc_k(\alpha)\,w^{-\alpha}\,\,\,,
\end{equation}
where 
\begin{equation}
n_k=\frac{4}{k(k+1)(k+2)}
\end{equation}
is the exact solution of the rate equations for the connectivity distribution 
in the case of the linear connection kernel \cite{k1}, and $c_k(\alpha)$ is 
the normalization constant, determined by the condition 
\begin{equation}
c_k(\alpha)\hspace{-3mm}\int_{\hspace{4mm}E_k}^{\hspace{7mm} \infty}x^{-\alpha}\,dx=
1\,\,\,.
\end{equation}
The scaling expression (16) refers to cases when a weight $w$ is assigned to each 
link with probability determined for all $k$ by the same tail contribution 
$\rho(w)$ to the distribution $\rho_k(w)$, i.e. $\rho_k(w)\simeq\rho(w)$ for 
$w\gg 1$, where 
\begin{equation}
\rho(w)=\lambda w^{-\alpha}\,\,\,
\end{equation}  
with $\lambda>0$ being independent of $w$ and equal for all $k$. Consequently, 
substituting (16) into (15), using Eq. (17) and employing the condition (18), 
one finds for asymptotically large $w$ that  
\begin{equation}
\bigg(1+\frac{\varepsilon_{k-1}}{E_{k-1}}\bigg)^{\alpha-1}=1+
\bigg(\frac{\varepsilon_{k-1}}{E_{k-1}}\bigg)^{\alpha-1}+
O\Big(\frac{1}{w}\Big)\,\,\,,
\end{equation}
where the relation $E_k=E_{k-1}+\varepsilon_{k-1}$ has been applied. Eq. (20) 
holds if $\alpha=2$ (for $w\rightarrow\infty$). Thus, in the case of 
$\alpha =2$, the scaling function (16) indeed satisfies Eq. (15) when $w$ 
tends to infinity. Accordingly, the total weight distribution 
$P(w)=\sum_{k}n_k(w)$ undergoes for asymptotically large $w$ the power-law
\begin{equation}
P(w)\simeq b\,w^{-2}\,\,\,,
\end{equation}
where the constant $b=\sum_kn_kc_k(\alpha)$. By means of (17) and (18), this 
constant is finite when $k^{-2}E_k\rightarrow 0$ as $k\rightarrow\infty$.
The weight distribution has also been shown to display a power-law behavior 
in cases of networks in which the growth is governed by homogeneous and 
inhomogeneous attachment rules depending linearly on the degree of the 
existing node that connects new node, and in which links receive weights 
being powers of products of degrees of joined nodes \cite{am}.  

The above analysis shows that, in networks with connectivity-driven topology 
and with a special distribution $\rho_k(x)$ of weights assigned to links, 
the rate equations for the time evolution of the distribution of weights 
concentrated at vertices with a given connectivity have, for asymptotically 
large weights, a solution of the scale-free form. This solution refers to a 
simple case when the numbers of vertices of different degrees are all 
subjected to the power-law for the weight with the same scaling exponent 
$\alpha$. Thus, the solution to the considered rate equations possesses the 
power-law form with the same scaling index also in cases when the preference 
rate of attaching vertices is not linear \cite{k1}, provided that the 
distribution of weights assigned to links is of the form (19). The question 
of whether the studied rate equations for the total weight distribution have 
also other solutions (in cases of other distributions $\rho_k(x)$ than the one 
considered above), for which the total weight distribution $P(w)$ would 
reveal a scaling form at large $w$, is open. Clearly, what values could take 
the scaling index describing the power-law decay of the total weight 
distribution, or whether this distribution could undergo the power-law at 
all, it would depend on specific rules of link loading in an individual 
network. Note that the rate equation approach has been applied to investigate 
weight and degree distributions in networks with weight-driven 
topologies \cite{an}.

\end{document}